\documentclass[tinyml]{acmart}

\AtBeginDocument{%
  \providecommand\BibTeX{{%
    \normalfont B\kern-0.5em{\scshape i\kern-0.25em b}\kern-0.8em\TeX}}}

\setcopyright{rightsretained}
\copyrightyear{2022}
\acmYear{2022}

\definecolor{mygreen}{rgb}{0, 0.55, 0.55} 
\definecolor{mylg}{rgb}{0.51, 0.76, 0.78} 
\definecolor{myred}{rgb}{0.92, 0.27, 0.27} 
\begin{document}

\title{Improving the Energy Efficiency and Robustness of tinyML Computer Vision using Log-Gradient Input Images}






\author{Qianyun Lu}
\affiliation{%
  \institution{Stanford University}
  \city{Stanford}
  \country{United States}}
\email{savylu@stanford.edu}

\author{Boris Murmann}
\affiliation{%
  \institution{Stanford University}
  \city{Stanford}
  \country{United States}}
\email{murmann@stanford.edu}


\renewcommand{\shortauthors}{ }
\begin{abstract}
This paper studies the merits of applying log-gradient input images to convolutional neural networks (CNNs) for tinyML computer vision (CV). We show that log gradients enable: (i) aggressive 1.5-bit quantization of first-layer inputs, (ii) potential CNN resource reductions, and (iii) inherent robustness to illumination changes (1.7\% accuracy loss across $2^{-5} \cdots 2^3$ brightness variation vs. up to 10\% for JPEG). We establish these results using the PASCAL RAW image data set and through a combination of experiments using neural architecture search and a fixed three-layer network. The latter reveal that training on log-gradient images leads to higher filter similarity, making the CNN more prunable. The combined benefits of aggressive first-layer quantization, CNN resource reductions, and operation without tight exposure control and image signal processing (ISP) are helpful for pushing tinyML CV toward its ultimate efficiency limits.
\end{abstract}







\maketitle

\section{Introduction}
\begin{figure}[t]
  \centering
  \vspace*{.66cm}
  \includegraphics[width=.95\linewidth]{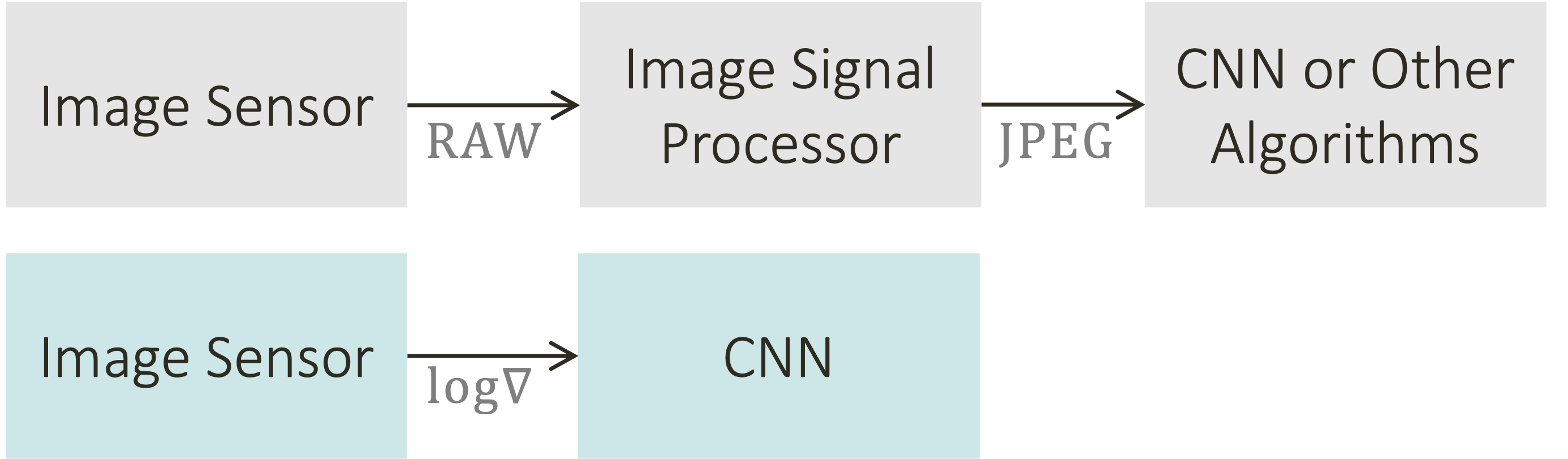}
  \caption{Conventional (top) and our $\log\nabla$ computer vision pipeline (bottom).}
  \label{fig:cvs}
  \Description{test test.}
\end{figure}

Fueled by advancements in machine learning algorithms and hardware, computer vision has extended its reach to battery powered Internet of Things (IoT) devices, enabling a variety of new applications that thrive on near-sensor data analytics \cite{shafique}. However, to further expand the commercial adoption of this technology, both the cost and energy consumption must continue to improve steadily. Our work focuses on this need through the lens of holistic hardware optimization, considering the computer vision (CV) pipeline from the physical image sensor to the classifier output.
As shown in Figure \ref{fig:cvs}, the traditional CV pipeline consists of three parts: the image sensor, the image signal processor (ISP), and task-specific algorithms such as convolutional neural networks (CNNs) \cite{euphrates}. The image sensor produces raw images that are not visually appealing (to humans) and hence the ISP performs a variety of operations such as white balancing and gamma correction. While prior work has shown that a subset of ISP operations can be omitted for machine consumption \cite{reconfig_cv_pl, isp4ml, minISP}, our approach omits the digital ISP block entirely and instead delegates a small amount of preprocessing to the analog circuitry of the image sensor. Specifically, we explore the use of log gradients ($\log\nabla$) as a direct input to CNNs. A key feature of $\log\nabla$ is that they convey a normalized, illumination-invariant representation of the characteristic edges in the input image. The feasibility and energy efficiency of extracting log gradients within the image sensor has already been established in \cite{chrisy}, but this investigation was based on algorithms outside the modern deep learning framework. Our main contribution is to examine the merits of log gradients for CNN-based systems.


\begin{figure}[t]
  \centering
  \includegraphics[width=1\linewidth]{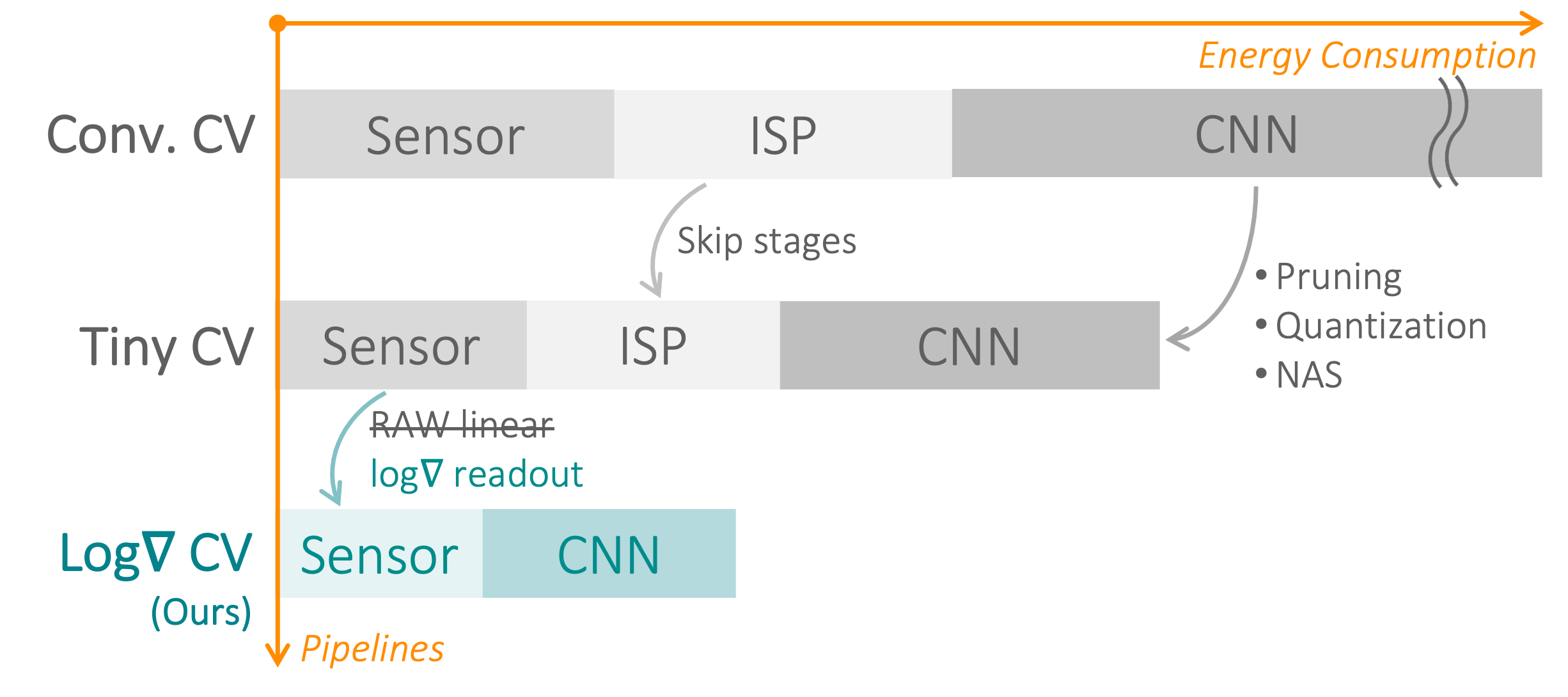}
  \caption{Energy breakdown of CV pipelines.}
  \label{fig:e_breakdown}
  \Description{test test.}
\end{figure}

Figure \ref{fig:e_breakdown} illustrates the motivation for our work from an energy perspective. In the shift from conventional CV pipelines to a solution optimized for tiny machine learning (tinyML), most energy savings have come from the remarkable progress in CNN compression \cite{model_compress}. However, as these compression techniques continue to yield smaller CNN footprints, the rest of the pipeline becomes important and can no longer be ignored. For context, we note that optimized hardware for small CV CNNs operates at around 3-6 nJ/pixel \cite{eb_3nj2, eb_6nj7}. This number is expected to decrease significantly given the large community working on CNN optimizations across the software-algorithm-hardware stack \cite{shafique}. Depending on their specs, image sensors consume about 0.2-2 nJ/pixel \cite{eb_0nj2}, while high-end ISP macros consume 1-2 nJ/pixel \cite{eb_1nj} in the latest CMOS technology. While skipping ISP stages can reduce the latter significantly, we show that the benefits of a log-gradient approach is not merely limited to ISP energy savings but extends to potential improvements in robustness and CNN compressibility.

The following sections expand on the contributions of this work, which can be summarized as follows:

\begin{itemize}

\item We formulate the concept of employing a $\log\nabla$ sensor readout within a tinyML CV pipeline and discuss data set aspects.

\item We use neural network architecture search (NAS) to show that a CNN with $\log\nabla$ inputs becomes more compressible (Section 3.1). We offer some intuition for this behavior using filter similarity and validate this hypothesis using experiments with a fixed 3-layer toy CNN (Section 3.2).

\item We show that the $\log\nabla$ approach enables aggressive quantization of the CNN’s first layer inputs down to 1.5 bits (3 levels). This not only reduces data movement between the image sensor and the CNN, but also helps with memory requirements. The quantization of CNN weights and activations is not considered here and remains as future work.

\item We perform experiments to assess the robustness to simulated illumination changes for $\log\nabla$, RAW and JPEG inputs. We observe only 1.7\% accuracy loss for $\log\nabla$ with 8x change in brightness versus up to 10\% for JPEG. Robustness aspects are often neglected in the CNN compression literature but are becoming increasingly important as real-word applications proliferate.

\end{itemize}

\section{Log $\nabla$ CV pipeline}

\subsection{Log Gradient Computation}
Denote an image by $P\in \mathbb{R}^{H\times W}$, where $H$ and $W$ are the image height and width in pixels. We compute the $\log\nabla$ of an image as:
\begin{equation}
    P' = \log (P+1)
\end{equation}
\begin{equation}
    \log\nabla = P'*f \quad \quad \text{where} \quad
    f = \begin{bmatrix} 0 & -1 & 0 \\-1 & 0 & 1 \\ 0 & 1 & 0 \end{bmatrix}
\end{equation}
where the ``$+1$'' is needed to shift the input into the domain of the $\log$, ``$*$'' means 2D-convolution, and $f$ is a gradient filter that extracts and combines horizontal and vertical gradients. For RGB data, the gradients can be computed channel-wise, but we consider only grayscale images in the present work. As shown in \cite{chrisy}, the gradient computations can be performed efficiently within the analog readout circuitry of an otherwise standard image sensor. Though this aspect is not the focus of this paper, we provide a brief review here. Consider for example the difference between two logarithms. We can write:
\begin{equation}
    d = \log(a)-\log(b) = \log(a/b) \approx Q(a/b)
\end{equation}
where $Q$ represents a quantizer with log-spaced decision levels. The work of \cite{chrisy} thus uses a ratio-to-digital converter (RDC) instead of a conventional analog-to-digital converter (ADC). 

An important aspect that we will examine below is that the gradients (and hence the ratios) can be coarsely quantized, making the log distortion of only a few decision levels a relatively simple task. In absence of the specialized RDC readout used in \cite{chrisy}, it can be emulated in hardware using a standard image sensor readout with relatively simple, equivalent digital postprocessing operations (no logarithms needed).

\subsection{Log Gradient Intuition}
\begin{figure}[t]
  \vspace*{.05cm}
  \includegraphics[width=1\linewidth]{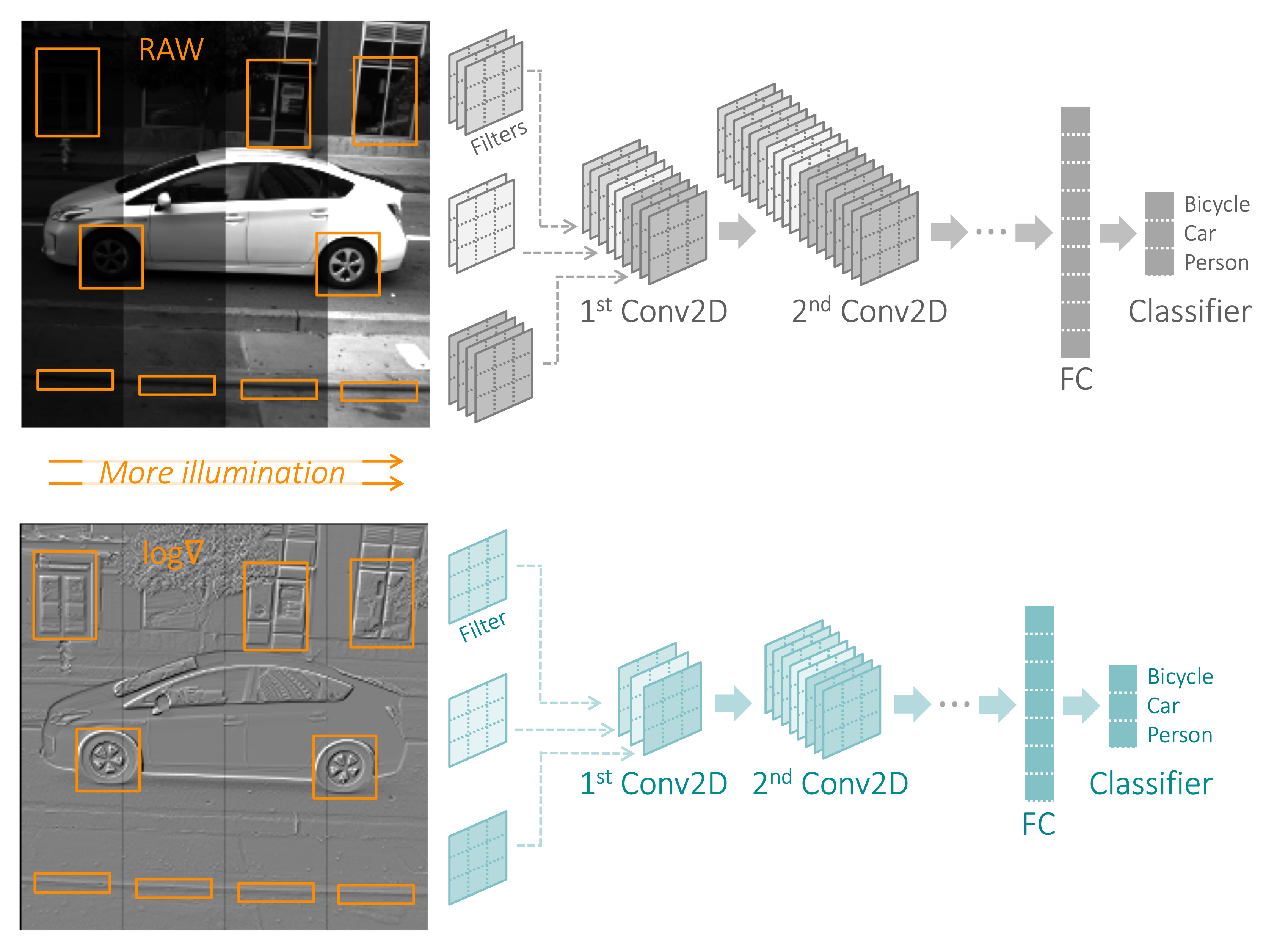}
  \caption{Illumination invariance of $\log\nabla$ and intuition about Conv2D filter savings.}
  \label{fig:illu_invar}
\end{figure}

To gain intuition about the potential benefits of log gradients, consider the contrived example in Figure \ref{fig:illu_invar}. The RAW image (top) contains four illumination levels while the computed $\log\nabla$ (bottom) image strips these away. The reason for this is that the difference of logarithms corresponds to ratios, and the pixel ratios in the top picture are not affected by the illumination changes. To keep this discussion concise, we ignore various second-order effects related to noise and saturation; the interested reader can refer to \cite{chrisy}.

Given the normalizing behavior of log gradients, we expect to see three benefits. First, while the $\log\nabla$ representation is not visually appealing, it should help in making the CNN classification more robust to global illumination changes. In a tinyML application, this may help relax exposure control requirements. Second, since the $\log\nabla$ operation reduces the image's dynamic range, it should be more amenable to aggressive quantization. Both of these properties have already been established in \cite{chrisy} but will be reiterated here for CNN-based applications. Third, since CNNs basically perform pattern matching using Conv2D filters, the RAW image should need a larger variety of (scaled) filters to detect the same object under different illuminations. In contrast, $\log\nabla$ should only need one filter for each characteristic pattern, hence enabling a smaller model size (see Figure \ref{fig:illu_invar}). 

\begin{figure}[t]
  \vspace*{.05cm}
  \includegraphics[width=1\linewidth]{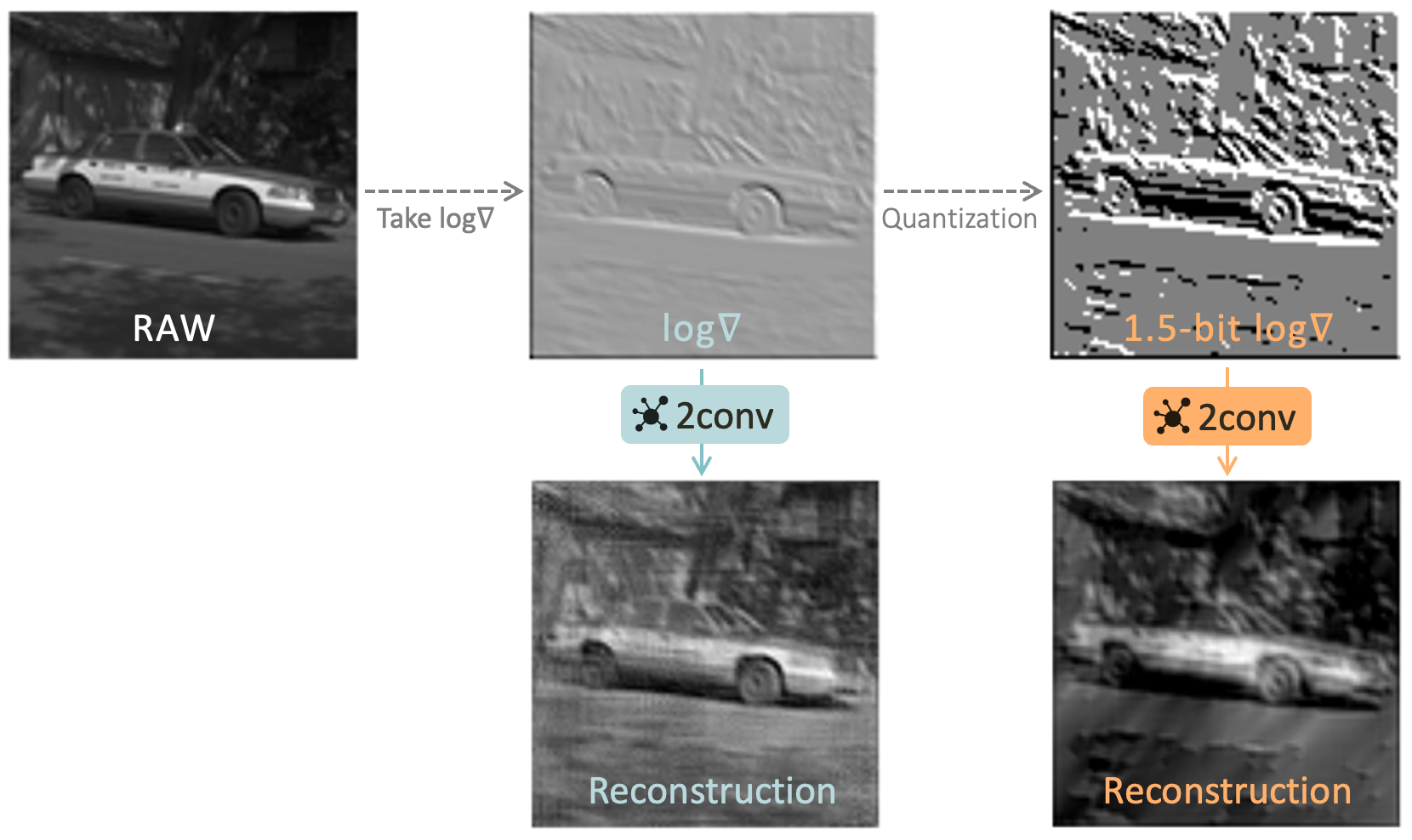}
  \caption{Image reconstruction from $\log\nabla$ using 2conv.}
  \label{fig:reconstructions}
\end{figure} 

To provide a feel for the losses introduced by the $\log\nabla$ representation, we reconstructed an original RAW image from its log gradients using a two-layer toy CNN (2conv). Figure \ref{fig:reconstructions} shows the outcomes without quantization and for quantized 1.5-bit (3-level) $\log\nabla$. Despite the smudges, we are able to recognize the car in both cases, indicating that the $\log\nabla$ preserve most of the useful information even with aggressive quantization. Note that this example is for illustrative purposes only (it would make no sense to reconstruct the original image in the end application). Details of the training and inference setups for this toy example are provided in Appendix \ref{apdx:2conv_recon}.

\label{sec:prop_cv}

\subsection{Datasets}
Experimentation with a $\log\nabla$ pipeline requires RAW image data. Unfortunately, most public datasets consist of post-ISP 8-bit JPEG images, which are obtained through irreversible and lossy transformations. An exception is the PASCAL RAW 2014 dataset \cite{raw2014}, which contains physical image sensor data for three object types: bicycle, car, and person. Our experiments in Section 3 use images from this dataset that were cropped according to classification bounding boxes. This set includes 6550 images total (bicycle : car : person = 708 : 1765 : 4077). All images are 16-bit grayscale after demosaicing. To prove that our approach is not overly sensitive to the choice of dataset, we performed ancillary experiments using approximate RAW data reconstructed from the popular Visual Wake Words (VWW) dataset (see Appendix \ref{apdx:vww}).





\section{Experiments}
To demonstrate the advantages of the $\log\nabla$ approach, we performed experiments using several formats for the first-layer CNN-inputs:
\begin{eqnarray*}
    \mathcal{I} = \{\hspace{.3cm}\text{8-bit JPEG}, & \text{16-bit RAW},  \\
    \text{FP $\log\nabla$},  & \text{1.5-bit $\log\nabla$}, & \text{2.25-bit $\log\nabla$} \hspace{.3cm}
    \}.
\end{eqnarray*}
Here, 16-bit RAW corresponds to demosaiced grayscale images from the PASCAL RAW 2014 dataset. Full precision floating point (FP) $\log\nabla$ are log gradients computed from these raw data without explicit quantization. The labels 1.5-bit and 2.25-bit denote 3- and 5-level representations that were quantized using empirical thresholds (the threshold values are not critically sensitive, but can potentially be tuned during training). We generated the data labelled "8-bit JPEG" from the demosaiced RAW images using only gamma correction ($\gamma=2.2$). We found that using additional ISP stages made only insignificant differences in our experiments with small CNNs, which is in line with the conclusions of \cite{reconfig_cv_pl}. All experiments focus on image classification which can be further extended to downstream tasks like transfer learning and object detection.


Due to the lack of explainability of ML algorithms, we study the impact of $\log\nabla$ inputs from two different perspectives. First, we use NAS to show that $\log\nabla$ lead to lower CNN resource requirements. Next, we study the impact of $\log\nabla$ on a 3-layer toy network. We observe higher filter similarity and thus higher prunability with $\log\nabla$ inputs, which is consistent with the NAS results. 

\subsection{CNN architecture search}
\label{sec:unas_results}
We employ the $\mu$NAS algorithm \cite{unas}, which considers the available size of RAM, persistent storage and processor speed to derive bounds on peak memory usage, model size, and latency (approximated by the number of multiply-accumulate operations (MACs)). Aging evolution is used as the main search algorithm and can be combined with structural pruning. The search space has high granularity and few restrictions on layer connectivity. \textcolor{black}{ For details of uNAS' search space, refer to Table 1 in \cite{unas}.} All CNN-internal weights and activations are kept in floating point precision to cleanly isolate the impact of varying the input data format. Further details of the training setup are provided in Appendix \ref{apdx: unas}. 


Figure \ref{fig:unas_pareto_A} shows the resource versus error rate Pareto fronts when $\mu$NAS is fed with three different image types. Relative to the 8-bit JPEG baseline case, we see that going to 16-bit RAW leads to a significant increase in the required resources for a given error rate. This is consistent with the benefits of ISP observed in \cite{isp4ml}. However, when we input FP $\log\nabla$, we recover these losses and in fact outperform the JPEG baseline case. For example, for an error rate of 2\%, the model size, MACs and peak memory usage of FP $\log\nabla$ area about 5x, 5x, 1.1x smaller than for JPEG.

\begin{figure}[t]
  \centering
  \includegraphics[width=1\linewidth]{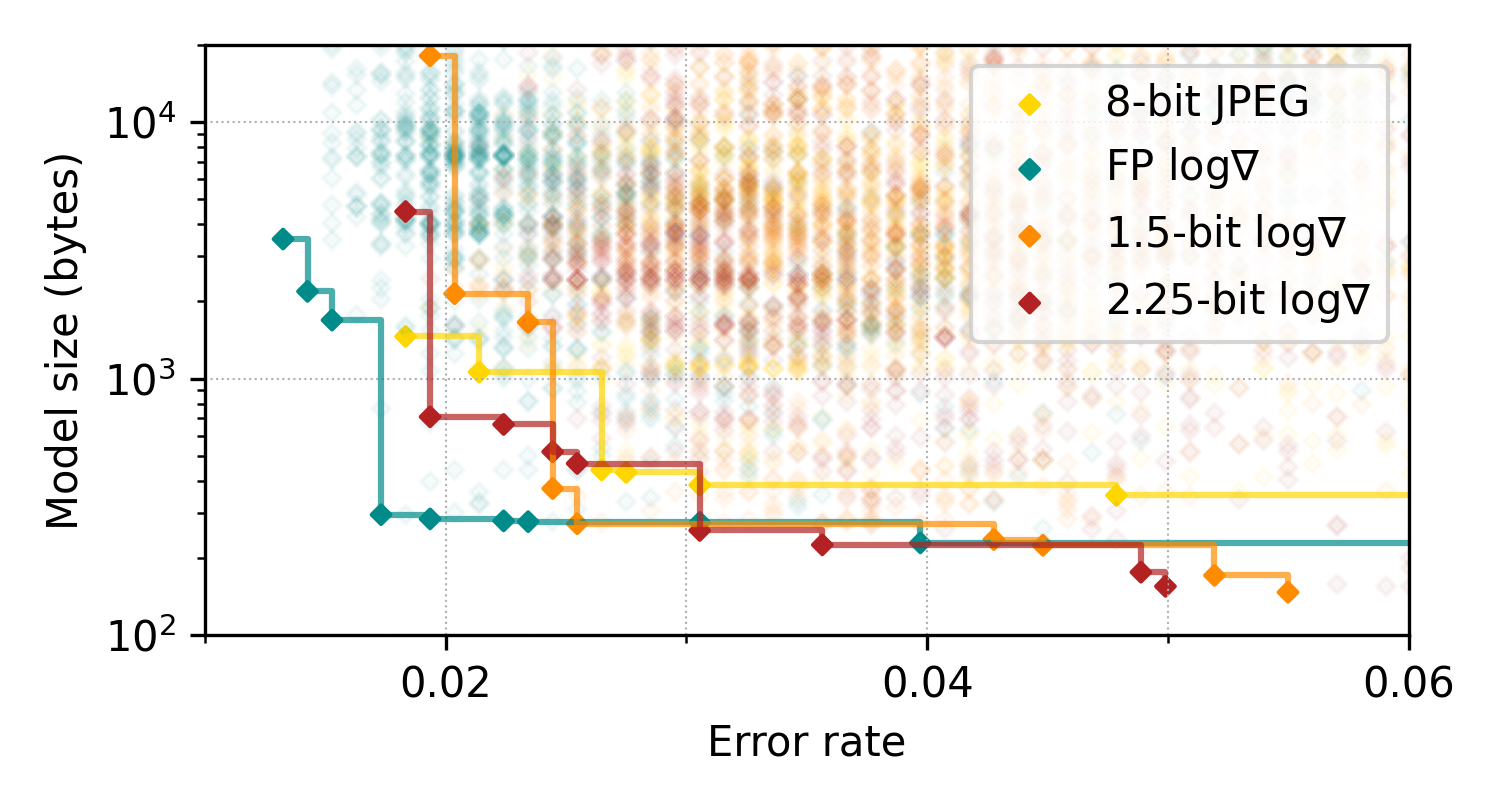}
  \vfill
  \includegraphics[width=1\linewidth]{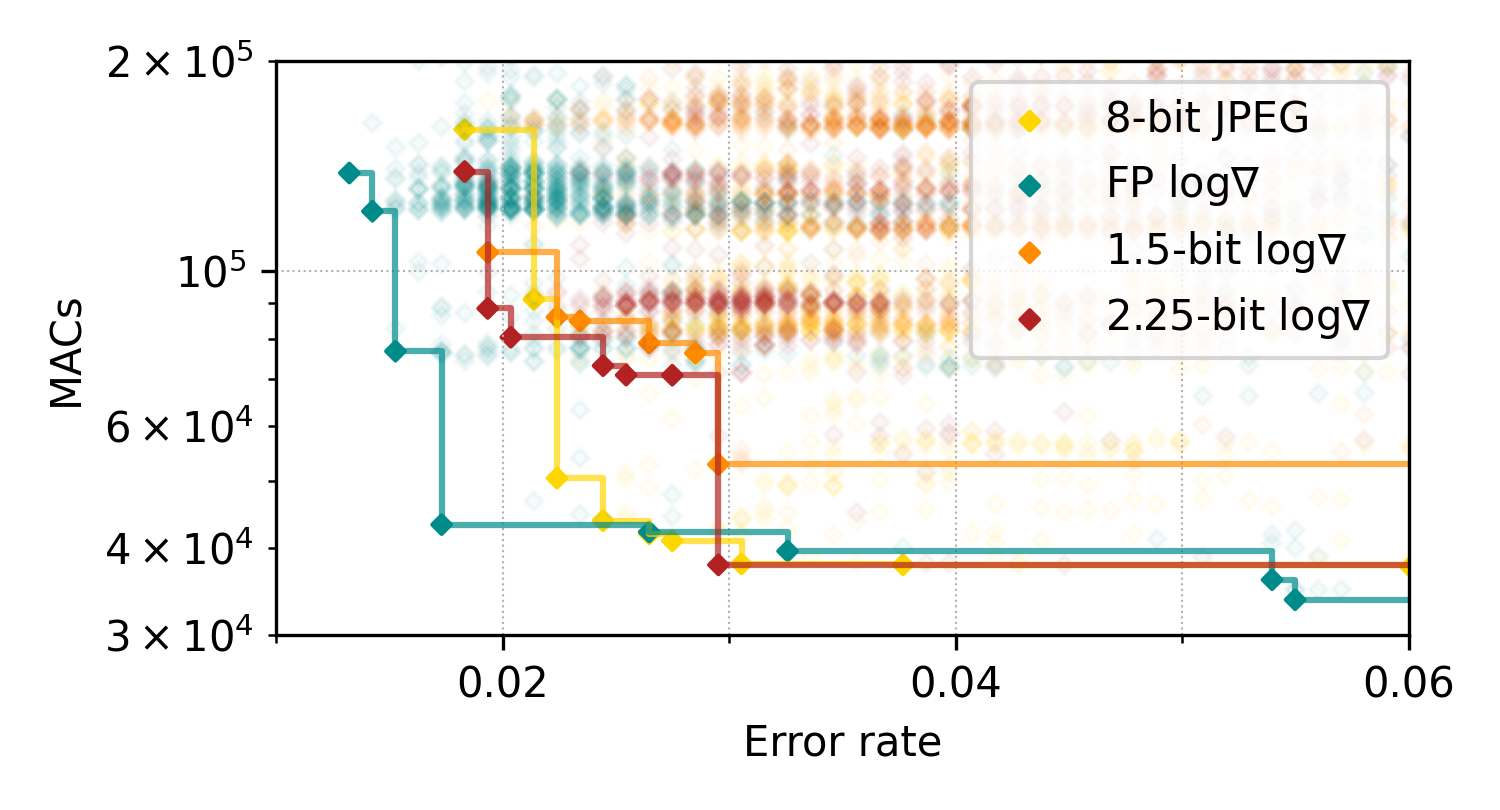}
  \vfill
  \includegraphics[width=1\linewidth]{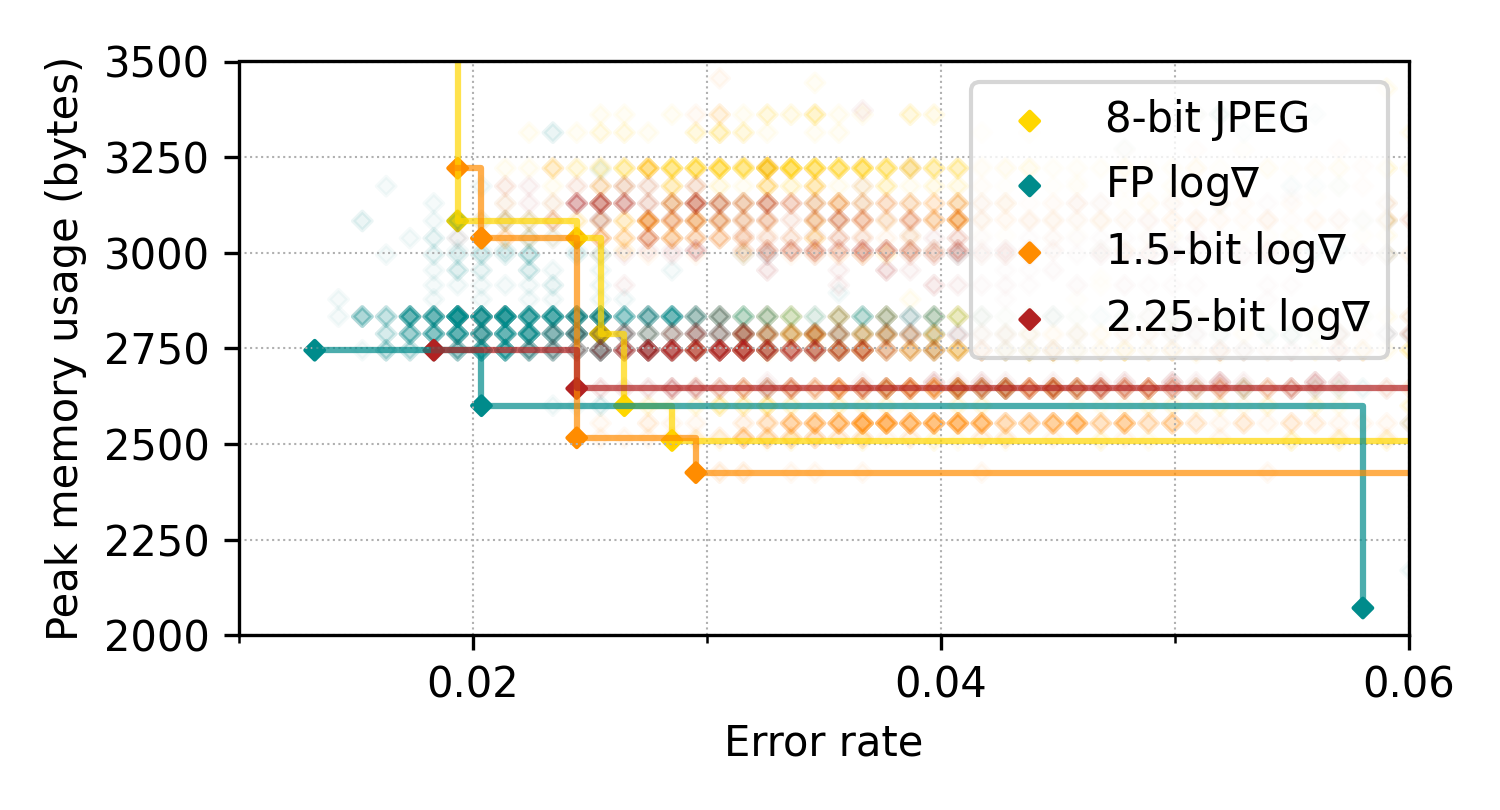}
  \caption{CNN resources vs. classification error rate for high-resolution and aggressively-quanitzed inputs.}
  \label{fig:unas_pareto_B}
\end{figure}

Since FP $\log\nabla$ would be too costly to compute in hardware, we repeat the experiment using the quantized 1.5-bit and 2.25-bit log gradient inputs (see Figure \ref{fig:unas_pareto_B}). As expected, there is some accuracy loss with quantization, but the degradations are modest and still in line with or slightly better than the JPEG baseline. We note that there is room for further experimentation with different hyperparameters as well as quantized CNN weights and activations for a more exact assessment of the required resources. The main conclusion from this experiment is that coarsely quanitzed $\log\nabla$ appear to be well suited as inputs to small, NAS-optimized CNNs. 

\begin{figure}[t]
  \centering
  \includegraphics[width=1\linewidth]{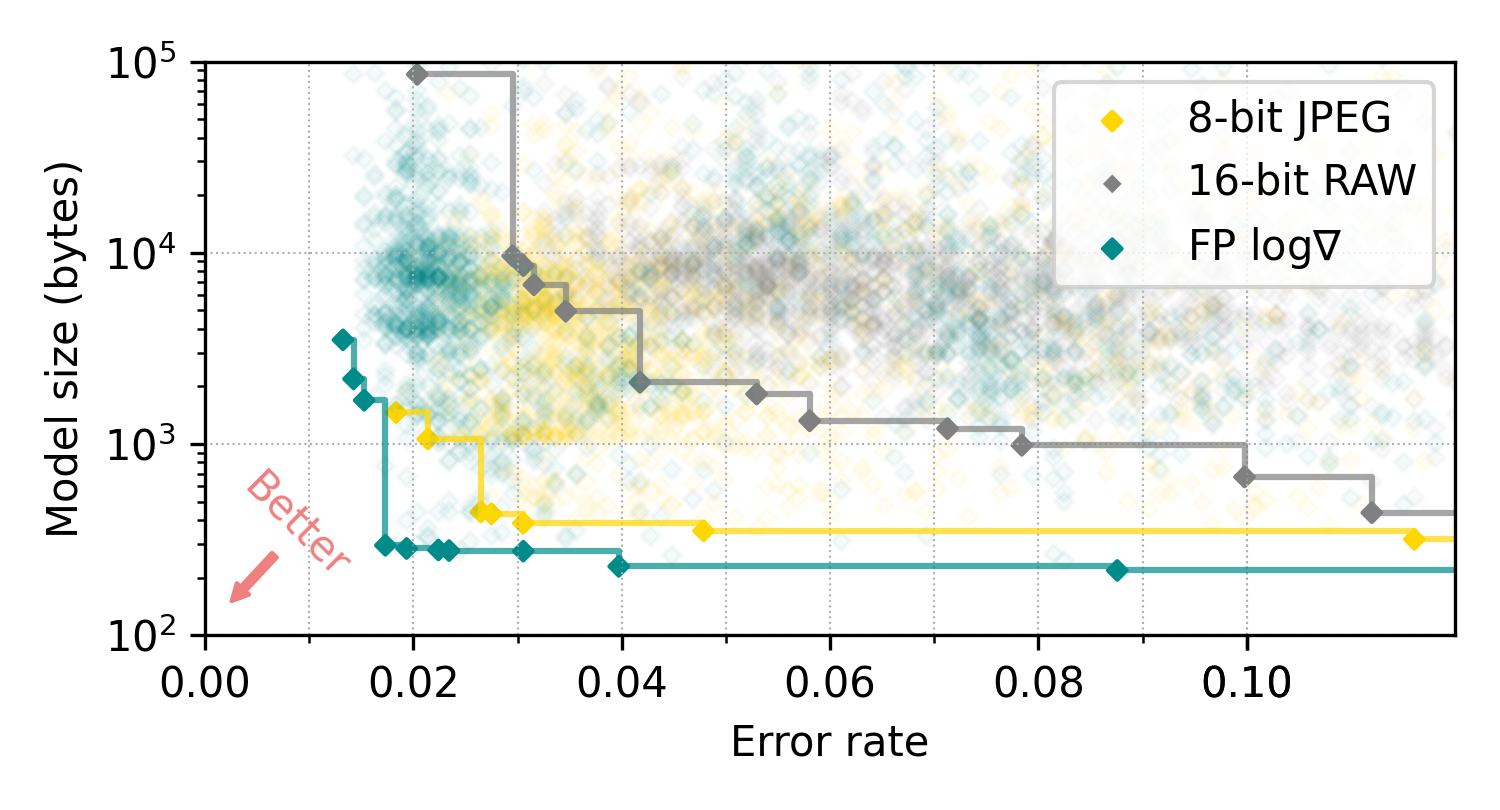}
  \vfill
  \includegraphics[width=1\linewidth]{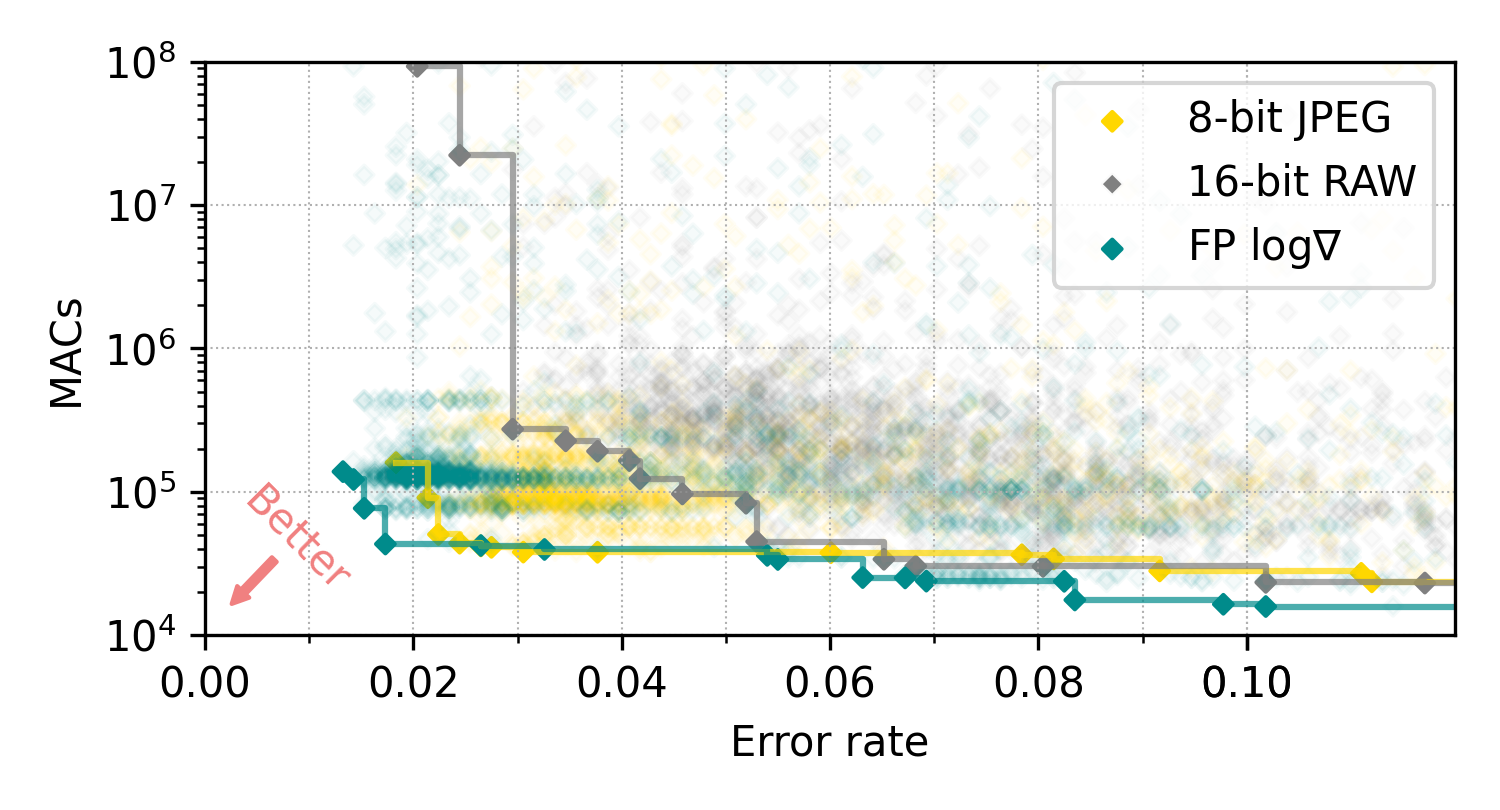}
  \vfill
  \includegraphics[width=1\linewidth]{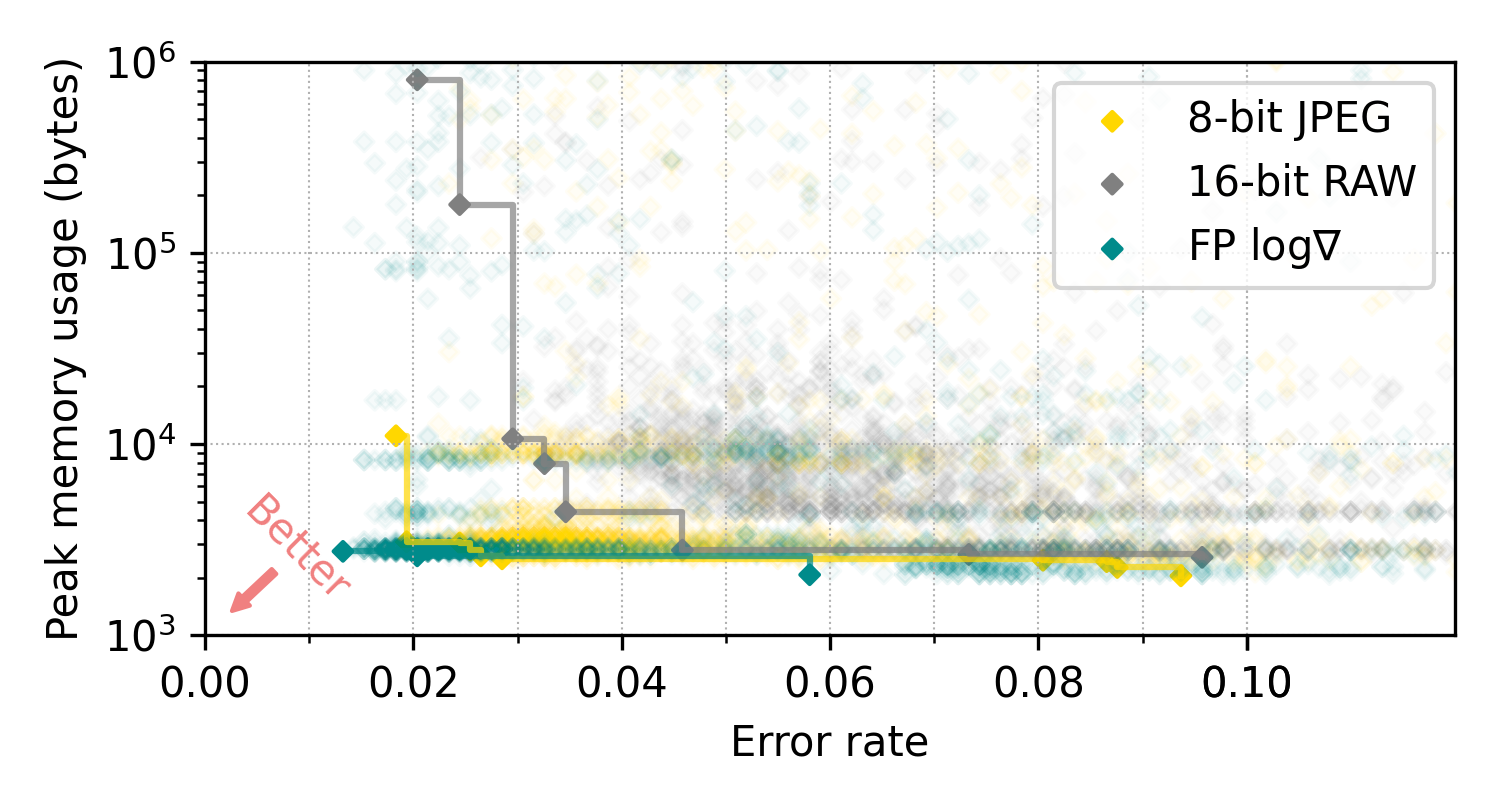}
  \caption{CNN resources vs. classification error rate for high-resolution inputs.}
  \label{fig:unas_pareto_A}
\end{figure}

\subsection{Fixed CNN architectures}
\begin{figure}[t]
  \centering
  \includegraphics[width=0.38\linewidth, angle=90]{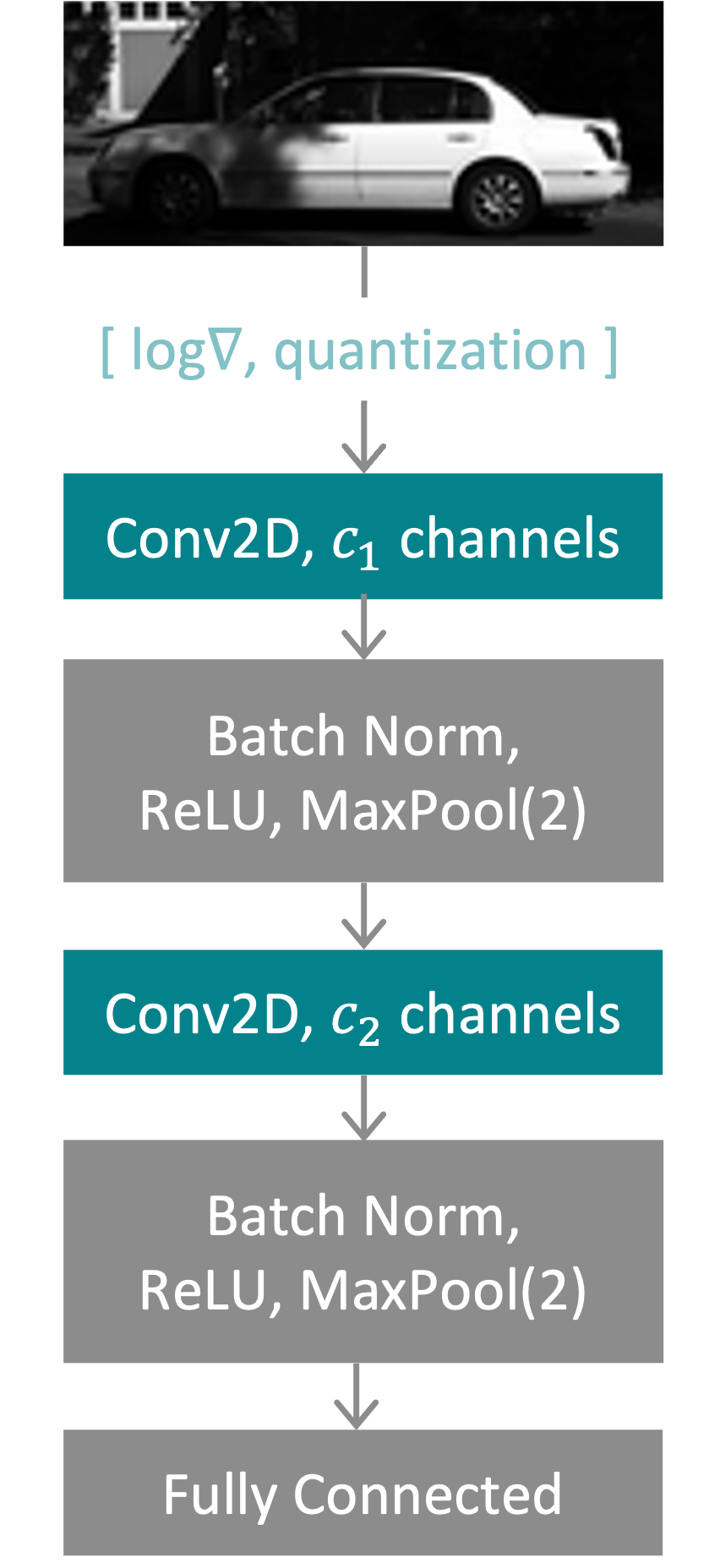}
  \caption{Toy CNN architecture 2conv1fc.}
  \label{fig:2conv1fc}
\end{figure}

Section \ref{sec:prop_cv} argued that $\log\nabla$ inputs should require fewer Conv2D filters than RAW data for classification. If this is true, we should observe a change in filter redundancy if we keep the CNN architecture fixed and vary the input $I\in \mathcal{I}$. Thus, the experiments in this section employ a toy CNN with two convolutional and one fully connected layer (2conv1fc, see Figure \ref{fig:2conv1fc}), using a kernel size of 5 and channel counts of $c_1=150$, $c_2=5$. Further details of the training setup are provided in Appendix \ref{apdx: 2conv1fc}. 

Higher filter redundancy means that there is a higher degree of similarity among the filters. To investigate, we thus compute the layer-wise cosine similarities among CNN filters \cite{duplicateFilters} after training:
\begin{equation}
    s_{i,j} = w'_i \cdot w'_j
\end{equation}
where $w'_i$ and $w'_j$ are normalized filters in the same Conv2D layer. 

\begin{figure}[t]
  \centering
  \includegraphics[width=.9\linewidth]{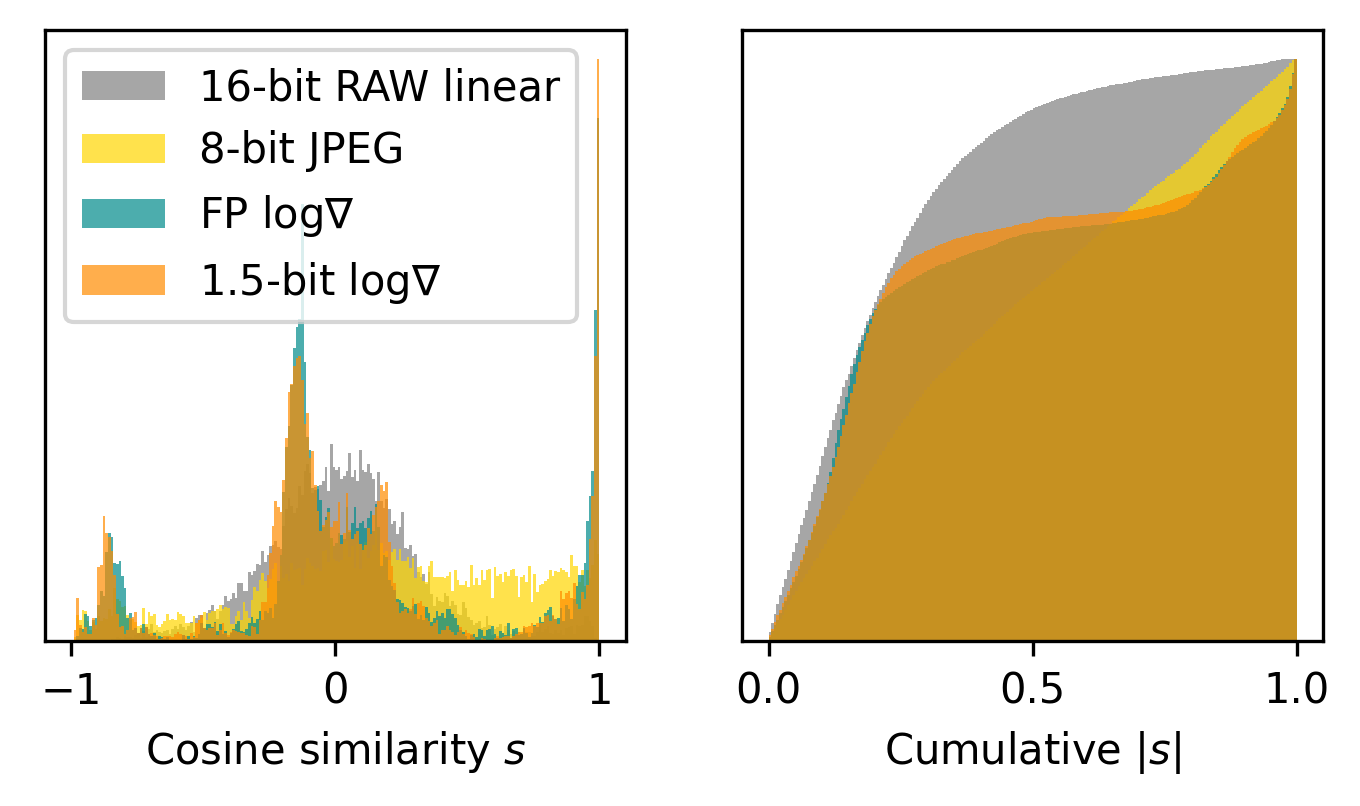}
  \caption{Histograms of cosine similarity.}
  \label{fig:cos_simi}
\end{figure}

Figure \ref{fig:cos_simi} shows histograms of filter similarities for a few representative cases. We observe that FP and 1.5-bit $\log$-$\nabla$ models have heavier tails around $s \approx 1$ when compared to the CNN trained on 16-bit RAW, implying that a larger fraction of the filters are similar. The same is true for JPEG, but to a lesser extent. For better interpretability, we also plot the cumulative histograms of absolute similarities $|s|$ which saturate abruptly as $s$ approaches 1 for the cases of FP and 1.5-bit $\log\nabla$, corresponding to the peaks at $s \approx 1$ in the normal histograms. Examples of highly similar filters are visualized in Appendix \ref{apdx:vis_filters}. 

\begin{figure}[t]
  \centering
  \includegraphics[width=.9\linewidth]{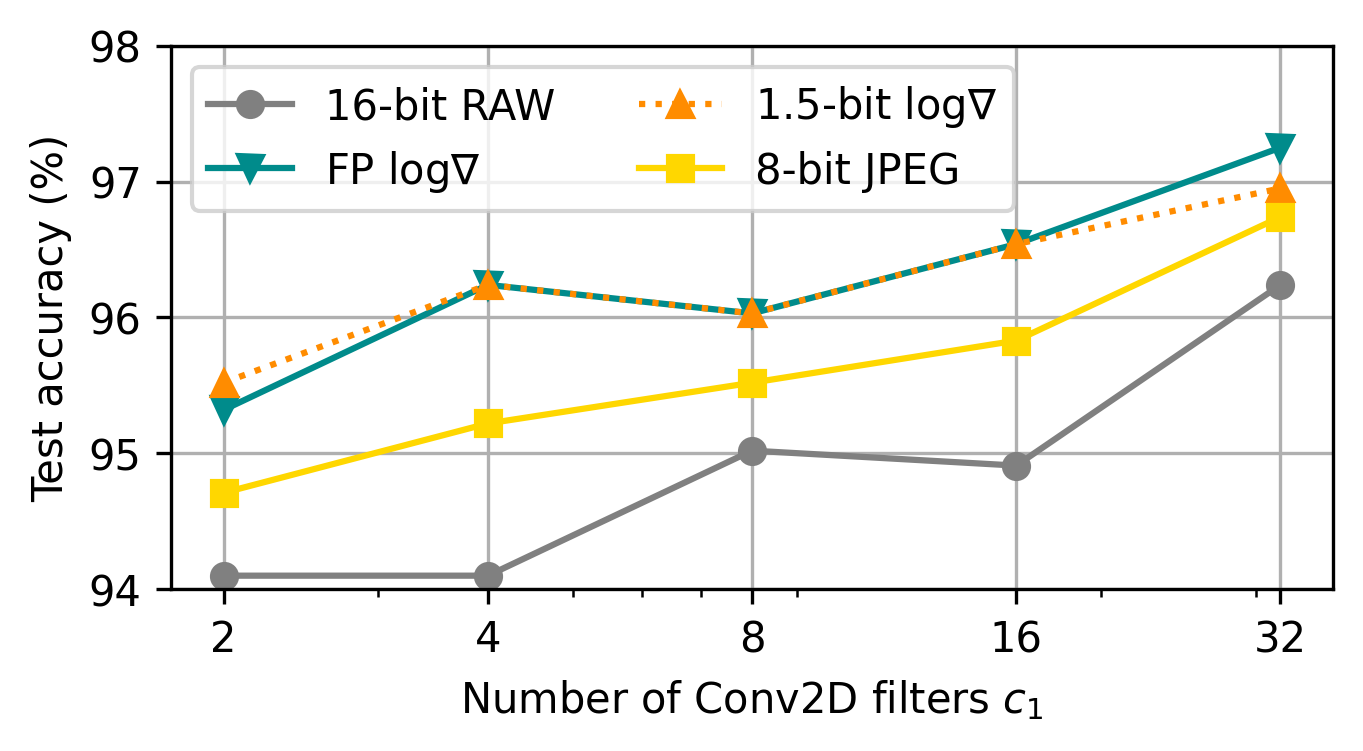}
  \caption{Test accuracy with $c_1\in \{2, 4, 8, 16, 32\}, c_2=8$.}
  \label{fig:fix_c2}
\end{figure}

RoyChowdhury \textit{et al.} have shown that higher filter similarity allows more channel pruning \cite{duplicateFilters}. We confirm this by fixing $c_2=8$ and varying $c_1\in \{2, 4, 8, 16, 32\}$. From Figure \ref{fig:fix_c2}, we see that $\log\nabla$ inputs need only 2 filters in the first Conv2D layer to achieve 95.3\% accuracy, whereas 16-bit RAW requires 32 filters (and JPEG performing somewhere inbetween). This further corroborates the above findings on filter redundancy. Using $\log\nabla$ gradient inputs induces more redundancy and hence leads to higher CNN prunability when compared to RAW and JPEG inputs.

\begin{figure}[t]
  \centering
  \includegraphics[width=.9\linewidth]{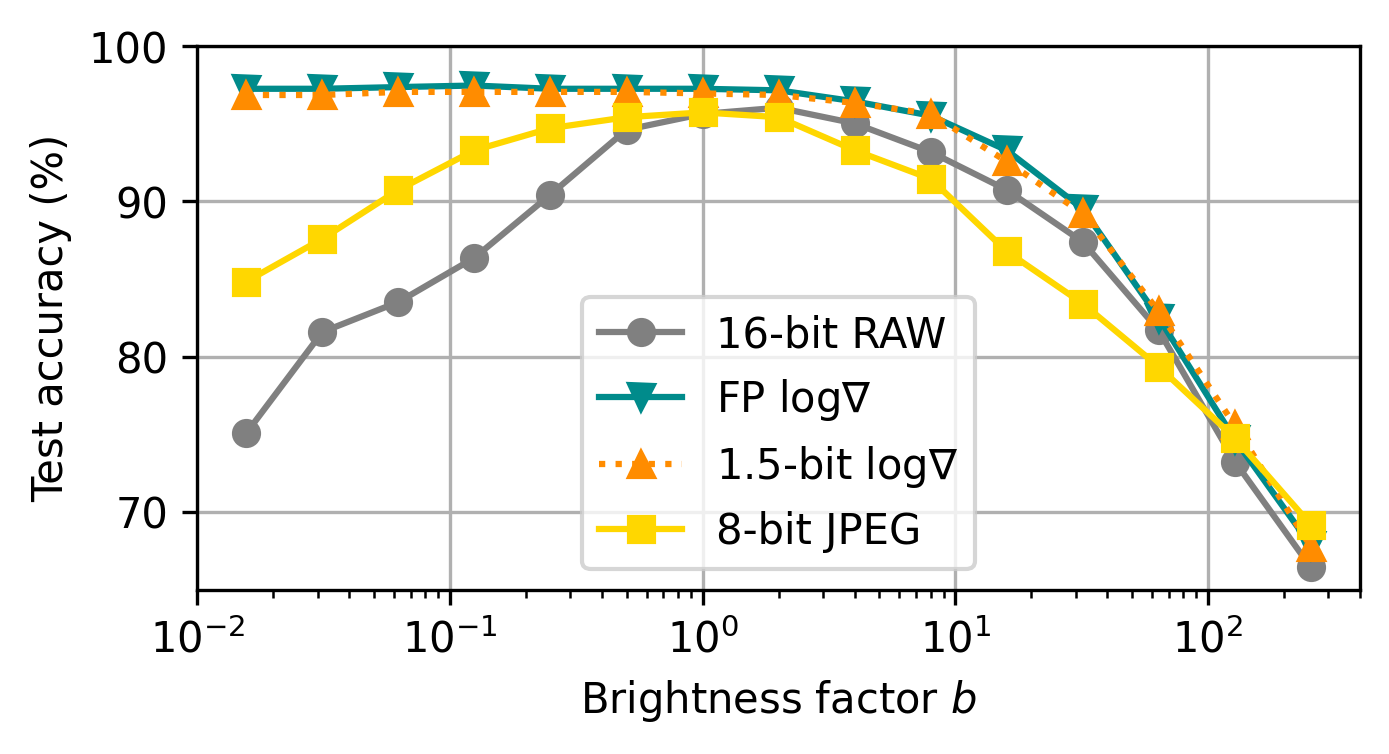}
  \caption{Test accuracy vs. image brightness.}
  \label{fig:sweep_brightness}
\end{figure}

\label{sec:illum}
As a final experiment using the fixed CNN, we consider the sensitivity to simulated illumination changes. We take the largest networks from the previous experiment ($c_1=32, c_2=8$) and vary the brightness of the test images by a factor $b \in \{2^{-6}, 2^{-5}, \cdots, 2^8\}$ relative to the nominal training brightness. The results in Figure \ref{fig:sweep_brightness} show that both FP and 1.5-bit $\log\nabla$ always have the best accuracy and most gradual drops when the images get brighter ($b>1$). Consistent with the results of \cite{chrisy}, the $\log\nabla$ accuracy also shows insignificant drops for darker images ($b<1$). In contrast, the 8-bit JPEG case shows significant drops in either direction, e.g. 10.9\% and 4.3\% accuracy deterioration with $2^{-6}\approx$0.016x and 8x brightness perturbations while the $\log\nabla$ models only drop by about 0.1\% and 1.7\%. This shows that gamma correction is a useful normalization, but it appears to be inferior to the the log-gradient approach. 

\section{Related Work}
\label{sec: related_work}
{\bfseries Model compression.} There exists a considerable body of literature on reducing the footprint of CNNs using compression techniques like weight and channel pruning as well as quantization \cite{model_compress}. A related approach that is also leveraged in our work is neural architecture search. For tinyML applications, NAS has been particularly effective in mapping CNN workloads onto resource-constrained microcontrollers \cite{unas, micronets, mcunet}. Our approach is complementary to this prior work since it focuses more on the input image representation. We have shown that log gradient inputs make the CNN more compressible and less sensitive to illumination. Additionally, our approach enables aggressive quantization of the first layer inputs down to 1.5 bits. Typically, at least 5 bits tend to be required using standard JPEG-based inputs \cite{fqconv}. In essence, our work focuses on stripping largely irrelevant information from the CNN input to enable better model compression. Recent work on RNN-based pooling for image patch summarization \cite{rnnpool} follows a related train of thought, albeit using more complex methods.

{\bfseries Dataset limitations.} Applications of tinyML CV will typically involve real-time image analysis ``in the wild,'' with varying camera angles and non-ideal or non-uniform illumination. Recent work has established that these effects can lead to significant accuracy reductions for CNNs trained on idealized data \cite{hdr4cv, openloris}. Unfortunately, popular datasets such as ImageNet \cite{imagenet} or Visual Wake Words (VWW) \cite{vww} (used for tinyMLPerf benchmarking \cite{tinymlperf}) are based on Internet photos and do not necessarily reflect what real-world sensors see. Additionally, the original RAW output from the image sensor cannot be reproduced (only estimated) from these JPEG datasets. We sidestep these issues by using the PASCAL RAW 2014 \cite{raw2014} dataset, which contains unprocessed RAW images. Access to the RAW pixel data allowed us to perform simulated illumination changes as shown in Section 3.

{\bfseries ISP simplifications.} Various researchers have pointed out that the traditional ISP pipeline is overprovisioned for image consumption by machines \cite{reconfig_cv_pl, isp4ml, minISP}. Buckler \textit{et al.} \cite{reconfig_cv_pl} found that demosaicing and gamma correction are the most critical ISP stages for small CNNs. Hansen \textit{et al.} \cite{isp4ml} performed an ablation study on ISP stages using the ImageNet dataset \cite{imagenet} and found that tone mapping is important for high dynamic range inputs. Overall, they argue that the energy consumed by some ISP operations is worth investing, since it takes a significantly larger neural network to achieve the same accuracy and generalization without preprocessing. The latter observation agrees with our results in Section 3, but we advocate for $\log\nabla$ as a simplified and tinyML-specific preprocessing alternative. Using $\log\nabla$ is motivated by physics and collapses the image’s dynamic range close to the sensor (essentially providing a more comprehensible alternative to gamma compression). This stands in contrast to traditional ISP operations, which are remnants of a pipeline optimized for image consumption by humans.

{\bfseries Logarithmic imaging.} The Weber-Fechner Law states that the intensity of human perception is proportional to the logarithm of the stimulus \cite{weber}, providing a biological motivation for log gradients. Log gradients have also been advocated in high-end imaging for motion rejection and noise reduction \cite{grad_cam}. The work of \cite{reconfig_cv_pl} also considered a logarithmic pixel readout but did not investigate gradient preprocessing. The work of Young \textit{et al.} introduced in-sensor log-gradient computation using a linear grayscale imager with ratio-to-digital conversion at 1.5 and 2.75 bits. The energy consumption of this image sensor is in line with the state of the art, consuming only 0.13 nJ/pixel with moderate dynamic range performance (59 dB). However, their backend was based on histograms of oriented gradients (HOG) with a deformable parts model (DPM) as the detection algorithm. These techniques are now outdated and hence this paper investigates $\log\nabla$ as inputs to modern CNNs for tinyML.

\section{Conclusion}
Image preprocessing plays an important role in computer vision but is often neglected in the literature. The de-facto standard in today’s ML research is to work with JPEG images that are appealing to the human eye. While it is well understood that some of the JPEG transformations are also beneficial for machine vision, tinyML systems of the future will look for an optimum solution that maximizes CNN compressibility and robustness. Our work identified $\log\nabla$ preprocessing as a promising option, as it enables aggressive quantization of first-layer inputs, improved CNN prunability, and increased robustness to illumination changes. Future work should consider training-based optimization of the $\log\nabla$ quantization thresholds, quantized training to reduce the internal compute precision of the CNN, as well as the response to adversarial inputs.


\begin{acks}
This work was supported in part by the ACCESS AI Chip Center for Emerging Smart Systems, sponsored by InnoHK funding, Hong Kong SAR. Qianyun Lu was supported by Stanford Graduate Fellowship in Science \& Engineering.
\end{acks}

\bibliographystyle{ACM-Reference-Format}
\bibliography{tinyml22lu_revised}

\appendix

\section{Image reconstruction from $\log\nabla$}
\label{apdx:2conv_recon}
\begin{figure}[t]
  \centering
  \includegraphics[width=.8\linewidth]{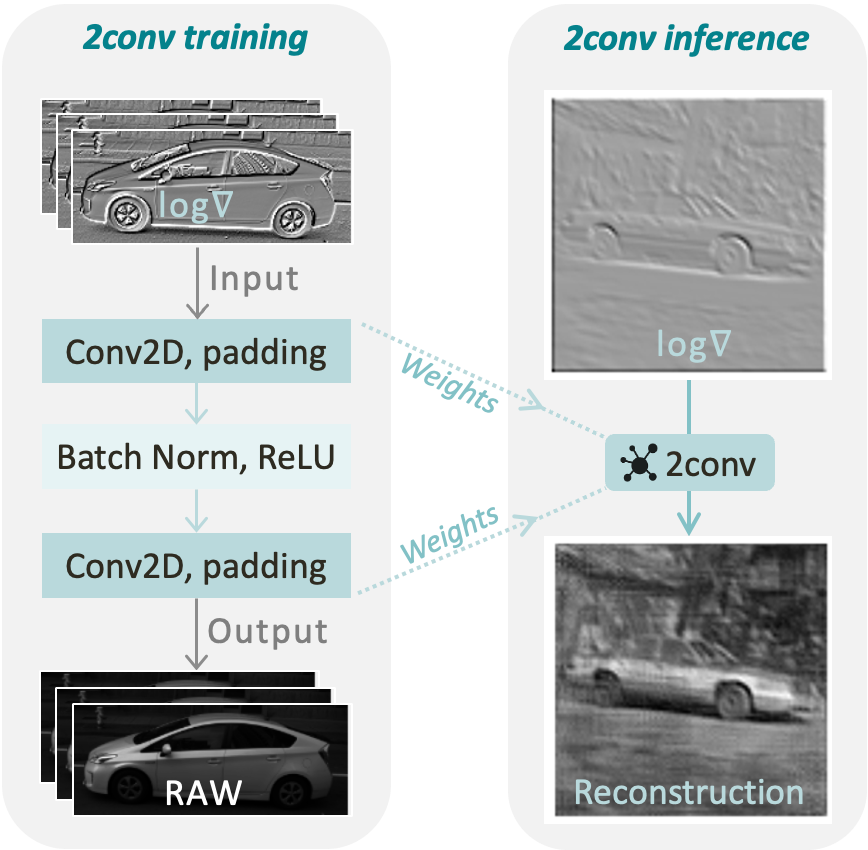}
  \caption{2conv training and inference for reconstruction.}
  \label{fig:2conv_recon}
\end{figure}

To reconstruct a RAW image from its $\log\nabla$, we train a two-layer network (2conv, see Figure \ref{fig:2conv_recon}) in TensorFlow, employing padding to ensure that the feature maps after the Conv2D layers have the same dimensions as the original input images. In the Conv2D layers, the channel counts are 10 and 1 respectively; filter sizes are chosen to be large to learn the global correlations in each training image (kernel size of 16 for 96x96x1 images). The loss is set as the mean square error between the RAW prediction and the ground truth (the original RAW images). The Adam optimizer is used with learning rate=0.0002. The batch size is 128 and the total number of epochs is 150 with the subset split of train : validation : test = 70 : 15 : 15 (\%).

\section{VWW+ExpandNet experiments}
\label{apdx:vww}
To approximate RAW data, we convert VWW JPEGs back to high dynamic range RGB images with the pretrained CNNs from \cite{expandnet}; the simulated RAW dataset is denoted by ``VWW+ExpandNet.'' Note that, even though the conversion yields higher dynamic range, it cannot recover all the information lost during JPEG compression. In other words, ``VWW+ExpandNet'' is just an approximation, not real RAW data. Images in the ``person'' class are cropped according to bounding boxes \cite{Zygarde} to avoid the random cropping and resizing problems when  ``person'' occupies a small area in an image. In other words, the dataset used is not the same as other works like MicroNets \cite{micronets}, and yields higher accuracy.

\begin{figure}[t]
  \centering
  \includegraphics[width=.9\linewidth]{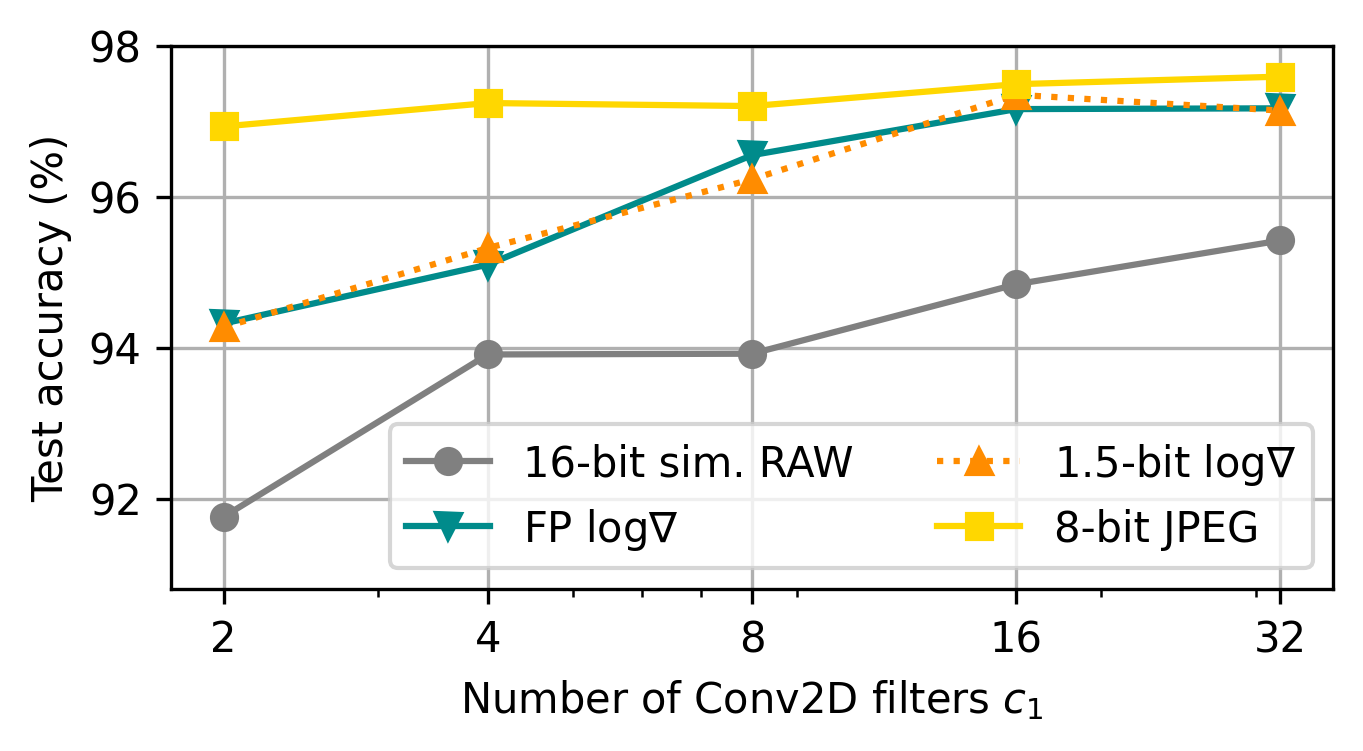}
  \caption{VWW test accuracy with $c_1\in \{2, 4, 8, 16, 32\}, c_2=8$.}
  \label{fig:vww_plot}
\end{figure}

Similar to the experiments in Section 3, we train the 2conv1fc CNN with $c_1\in \{2, 4, 8, 16, 32\}$, $c_2=8$ on four inputs $I \in \{$8-bit JPEG, 16-bit simulated RAW, FP $\log\nabla$, 1.5-bit $\log\nabla\}$, and plot the test accuracy in Figure \ref{fig:vww_plot}. The $\log\nabla$ models are less accurate than 8-bit JPEG models because ``VWW+ExpandNet'' is converted from non-invertible JPEGs, resulting in inaccurate $\log\nabla$ information. Nonetheless, the results are comparable for higher channel counts, indicating that our pipeline may generalize to other datasets. True RAW images would be required to come to a full conclusion. Compared with 16-bit simulated RAW, both FP and 1.5-bit $\log\nabla$ are more accurate with the same $c_1$ and need a smaller $c_1$ to achieve the same accuracy, which is consistent with our findings from Section 3.

\section{Training schedules}
The training details for the $\mu$NAS and the fixed CNN (2conv1fc) models are as given below. They were implemented in TensorFlow and PyTorch, respectively.

\subsection{$\mu$NAS training schedule}
\label{apdx: unas}
We train on images with dimensions 96x96x1 and the subset split is train : validation : test = 70 : 15 : 15 (\%). Aging evolution was used together with structured pruning. The batch size is 128 and the total number of epochs is 80 with pruning from 40\textsuperscript{th} epoch to 75\textsuperscript{th}; the minimum sparsity and maximum are set to 0.1 and 0.85, respectively. The TensorFlow SGDW optimizer is used with learning rate=0.03, momentum=0.9, and weight decay=0.0001. To limit the resources, the bounds are configured to: error bound=0.05, peak memory bound=10000, model size bound=20000, and mac bound=1000000. GPU info: 4x Nvidia GeForce rtx6000 @ 24GB GDDR6/module. 

\subsection{Fixed CNN training schedule}
\label{apdx: 2conv1fc}
We train on images with dimensions 224x224x1 and the subset split is train : validation : test = 70 : 15 : 15 (\%). The total number of epochs is 40 with a batch size of 64. The Adam optimizer is used with learning rate=0.001, scheduled to decay by gamma=0.95 every epoch. GPU info: 1x Nvidia GeForce rtx6000 @ 24GB GDDR6/module.




\begin{figure}[t]
  \centering
  \includegraphics[width=\linewidth]{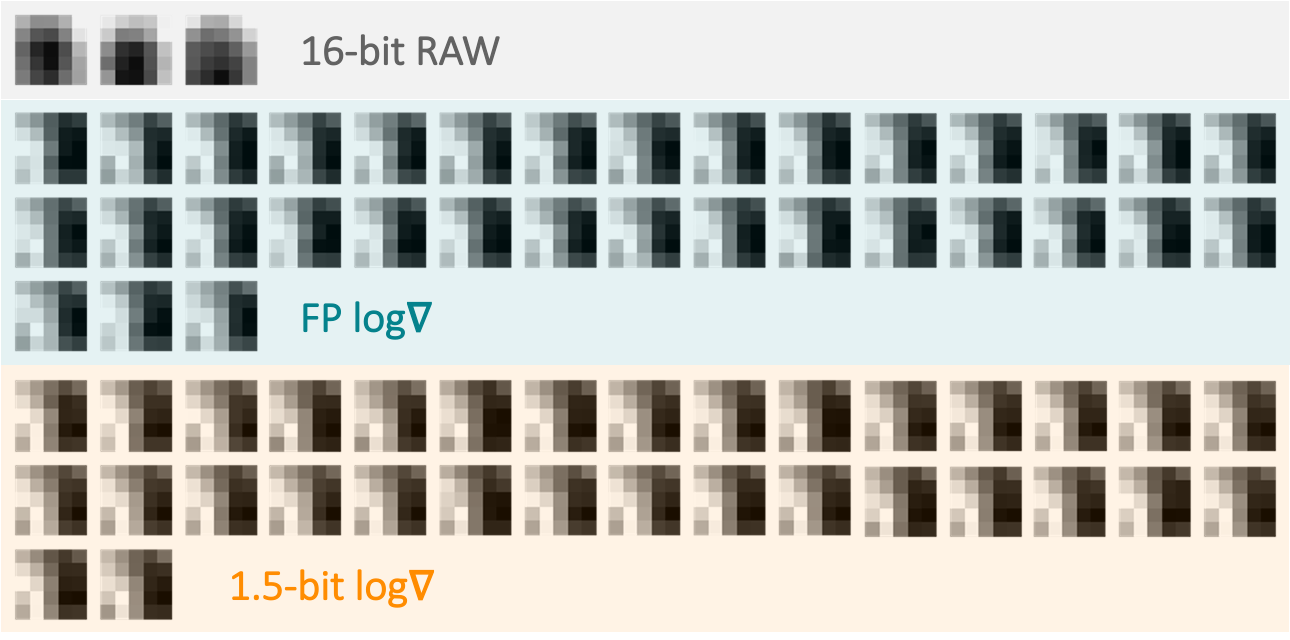}
  \caption{Visualization of similar filters.}
  \label{fig:visualize_filters}
\end{figure}

\section{Visualization of similar filters}
\label{apdx:vis_filters}
Figure \ref{fig:visualize_filters} shows filters of high similarity ($s_{i,j} > 0.98$). For both FP and 1.5-bit $\log\nabla$, we find a larger number of filters that are close to identical (about 11x more than for RAW).

\end{document}